\documentclass[12pt, a4paper]{article}
\usepackage{t1enc}
\usepackage[utf8]{inputenc}
\usepackage{a4}
\usepackage[dvips]{graphicx,rotating}
\usepackage{amsmath,amssymb,latexsym,mathrsfs}
\usepackage{subfigure}
\usepackage{color}
\usepackage{bbm}
\usepackage[bf,footnotesize]{caption}

\graphicspath{{plots/}}
\def \3{\ss }

\def\ppall{\mathaccent23p}

\newcommand{\beq}{\begin{equation}}
\newcommand{\eeq}{\end{equation}}
\newcommand{\beqn}{\begin{eqnarray}}
\newcommand{\eeqn}{\end{eqnarray}}

\newcommand{\be}{\begin{equation}}
\newcommand{\ee}{\end{equation}}
\newcommand{\bea}{\begin{eqnarray}}
\newcommand{\eea}{\end{eqnarray}}
\newcommand{\beaa}{\begin{eqnarray*}}
\newcommand{\eeaa}{\end{eqnarray*}}

\def\poz{a}
\def\hum{b}
\def\desy{c}
\def\liv{d}

\begin{document}
\begin{titlepage}
  {\vspace{-0.5cm} \normalsize
  \hfill \parbox{60mm}{DESY/08-014\\
                       HU-EP-08/02 \\
                       LTH 783 \\
                       SFB/CPP-08-15 \\}}\\[10mm]
  \begin{center}
    \begin{large}
      \textbf{Twisted Mass, Overlap and Creutz Fermions: \\
              Cut-off Effects at Tree-level of Perturbation Theory} \\
    \end{large}
  \end{center}

  \vskip 0.5cm
  \begin{figure}[h]
    \begin{center}
      \includegraphics[draft=false]{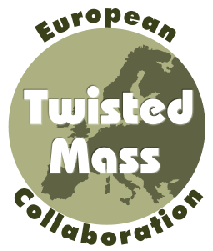}
    \end{center}
  \end{figure}

  \vspace{-0.8cm}
  \baselineskip 20pt plus 2pt minus 2pt
  \begin{center}
    \textbf{
      K. Cichy$^{(\poz)}$,
      J. Gonz\'alez L\'opez$^{(\hum,\desy)}$,
      K.~Jansen$^{(\desy)}$, \\
      A. Kujawa$^{(\poz)}$,
      A.~Shindler$^{(\liv)}$
      }
  \end{center}

  \begin{center}
    \begin{footnotesize}
      \noindent

$^{(\poz)}$ Adam Mickiewicz University of Poznan, Faculty of Physics,
Umultowska 85, \\ 61-614 Poznan, Poland\\
\vspace{0.2cm}

$^{(\hum)}$ Humboldt--Universit\"at zu Berlin, Institut f\"ur Physik,
Newtonstrasse 15, \\ 12489 Berlin, Germany\\
\vspace{0.2cm}

$^{(\desy)}$ DESY, Zeuthen, Platanenallee 6, D-15738 Zeuthen, Germany\\
\vspace{0.2cm}

$^{(\liv)}$ Theoretical Physics Division, Dept. of Mathematical Sciences,
\\University of Liverpool, Liverpool L69 7ZL, UK\\
\vspace{0.2cm}

    \end{footnotesize}
  \end{center}

\vspace{1.2cm}
  \begin{abstract}
    \noindent We study cutoff effects
at tree-level of perturbation theory
for maximally twisted mass Wilson, 
overlap and the recently proposed Creutz fermions. 
We demonstrate that all three kinds of lattice fermions exhibit 
the expected O($a^2$) scaling behaviour in the lattice spacing. 
In addition, the sizes of these cutoff effects are comparable 
for the three kinds of lattice fermions considered here. 
Furthermore, we analyze situations when twisted mass fermions are
not exactly at maximal twist and when overlap fermions are studied 
in comparison to twisted mass fermions when the quark masses are 
not matched.
  \end{abstract}
  
\end{titlepage}

\section{Introduction}

This paper is an investigation of the scaling behaviour towards the continuum limit
for different kinds of lattice fermions at tree-level of perturbation theory.
In a previous conference 
contribution \cite{Cichy:2007vk} we had only discussed the case
of Wilson twisted mass fermions
\cite{Frezzotti:2000nk,Frezzotti:2003ni}. 
See ref.~\cite{Shindler:2007vp} for a review on twisted mass lattice QCD.
Here we add overlap fermions \cite{Neuberger:1997fp}. See ref.~\cite{Niedermayer:1998bi} 
for a review on chirally symmetric lattice actions and related topics.
During the completion of this paper a new kind of lattice fermion appeared 
in the literature, the so-called Creutz fermions \cite{Creutz:2007af}
and we decided to investigate also these lattice fermions, 
as well as the fermions defined by a related action suggested 
by Borici \cite{Borici:2007kz}.
We will in the following generically refer to Creutz fermions 
having in mind both Creutz and Borici formulations. 

All of these three kinds of lattice fermions are expected to show an 
O($a^2$) scaling behaviour in the lattice spacing. 
While for overlap fermions this is achieved by an exact lattice
chiral symmetry, for twisted mass fermions this  
is achieved by a residual $N_{\rm f} = 2$ flavoured continuum chiral symmetry 
which needs, however, a tuning of the (untwisted) PCAC quark mass to zero.
Creutz fermions exhibit an exact $N_{\rm f} = 2$ flavoured continuum chiral
symmetry, but they break however a number of 
discrete symmetries such as parity, charge conjugation and time reflection
\cite{Bedaque:2008xs}. This could lead to a plethora of terms in the Symanzik
effective action, which would make approaching the continuum limit a rather difficult
task. 

Wilson twisted mass fermions are described 
by an ultra-local action with only nearest neighbour 
interactions. Hence, they are comparably cheap to simulate. 
Their major drawback is the explicit O($a^2$) isospin-breaking. 
Overlap fermions have the great advantage 
of an exact {\em lattice} chiral symmetry 
\cite{Luscher:1998pq}. However, while still local in the 
field theoretical sense \cite{Hernandez:1998et}, they couple each lattice 
point with all others and are an order of magnitude more 
expensive to simulate than twisted mass fermions. 
Finally, Creutz fermions \cite{Creutz:2007af} 
are also described by an ultra-local
action, again with only nearest neighbour interactions.                
As has been shown recently, they break however a number of 
discrete symmetries and isospin symmetry~\cite{Bedaque:2008xs}.

Twisted mass and Creutz fermions are defined
for $N_f=2$ flavours of quarks. These two quark flavours can either be 
taken as mass degenerate, or, some explicit flavour breaking term has to 
be added. In contrast, overlap fermions can be used for a single quark 
flavour.  

The focus in this paper is to study the scaling behaviour with the lattice spacing.
In particular, 
we are interested in the {\em relative} size of the 
cutoff effects comparing maximally twisted mass, overlap and 
Creutz lattice fermions.

In particular, we will consider the scaling behaviour of the 
meson correlation function
the pseudoscalar mass
and decay constant as well as the nucleon mass.
It is important to remark here that,
at tree-level of perturbation theory there is, of course, 
nothing like a real meson or baryon.
However, since we use the same interpolating fields as in full QCD,
we will take the freedom to use the notation of mesons and baryons
throughout this letter. 

We will also address the question of the size of cutoff effects 
when twisted mass fermions are not exactly tuned to maximal twist. In addition, 
we investigate ratios of mesonic quantities built from twisted and overlap
fermions when the quark masses are unmatched.

While actual practical calculations with overlap and twisted mass fermions are 
already rather advanced, the formulation of chiral invariant fermions
following Creutz is still very new. 

In \cite{Cichy:2007vk} we gave a first account
of the size of cut-off effects at tree-level of perturbation theory 
for twisted mass fermions. Here we extend the techniques described in 
\cite{Cichy:2007vk}, following \cite{Carpenter:1984dd}, to the case of overlap
and Creutz fermions. A more detailed account of our earlier calculations and 
a pedagogical introduction to the techniques we have used, can be
found in \cite{Thesis:2007}.

\section{Lattice propagators}

In this section we provide the momentum space propagators for the different
kinds of lattice fermions we have considered. They are the building blocks
for the computation of the meson and nucleon correlation functions from which in 
turn the masses and decay constants are extracted. 
To fix the notation our setup is a hypercubic lattice of size $L^3 \times T$ with spacing $a$.

\vspace*{0.6cm}
\noindent {\bf Wilson twisted mass fermions} 
\vspace*{0.6cm}

\noindent The expression for the Wilson twisted mass (Wtm) fermion propagator
in the twisted basis, at tree-level of perturbation theory (PT) and in momentum space 
can be derived from the Wtm operator given for example in ref.~\cite{Shindler:2007vp},
and it is given by
\begin{equation}\label{tmprop}
\widetilde{S}_{\rm Wtm}(p)= \frac{-i \ppall_\mu \gamma _{\mu}\mathbbm{1}_{f} +  M(p)
  \mathbbm{1}\mathbbm{1}_{f}-i\mu _{q}\, \gamma _{5}\tau _{3} }
{\sum_\mu \, \ppall_\mu^2 + M(p)^2 + \mu _{q}^{2}},
\end{equation}
where
\begin{equation}
\ppall_\mu = \frac{1}{a}\sin (ap_{\mu}) , \quad \hat{p}_\mu
=\frac{2}{a}\sin(\frac{ap_{\mu}}{2}) , \quad M(p) =  m_0 + \frac{a}{2}\,
\sum_{\mu}\, \hat{p}_\mu^2 ,
\end{equation}
and where $\mathbbm{1}$ and $\mathbbm{1}_{f}$ are the identity matrices in Dirac
and flavour space. The structure in colour space has not been written
since it is just an identity matrix at tree level of PT.
The parameters $m_0$ and $\mu_q$ represent the untwisted and twisted quark masses, 
respectively. Maximal twist --in the case of tree-level of perturbation theory--
is achieved by setting $m_0=0$. 
We then expect to have only O($a^2$) lattice 
spacing effects in physical correlation functions~\cite{Frezzotti:2003ni}.

\vspace*{0.6cm}
\noindent {\bf Overlap fermions} 
\vspace*{0.6cm}

The expression for the overlap propagator in momentum space at 
tree-level of perturbation theory can be derived from the expression of the overlap
operator given for example in ref.~\cite{Luscher:1998pq}, and it is
\begin{equation}
\tilde{S}_{\rm ov}(p)=\frac{-i(1-\frac{ma}{2})F(p)^{-1/2}\ppall_\mu \gamma_\mu + 
\mathcal{M}(p)\mathbbm{1}}{(1-\frac{ma}{2})^2 F(p)^{-1}\sum_\mu\ppall_\mu^2+\mathcal{M}(p)^2} 
\end{equation}
where:
\begin{equation}
 F(p)=1+\frac{a^4}{2}\sum_{\mu<\nu}\hat{p}_\mu^2 \hat{p}_\nu^2
\end{equation} 
\begin{equation}
\mathcal{M}(p)=\frac{1}{a}\Bigg(1+\frac{ma}{2}-\Big(1-\frac{ma}{2}\Big)
F(p)^{-1/2}\Big(1-\frac{a^2}{2}\sum_\mu \hat{p}_\mu^2\Big)\Bigg)
\end{equation} 
and $\mathbbm{1}$ is the identity matrix in Dirac space.
Note that in the case of overlap fermions we only discuss one flavour.
Due to the existence of an exact lattice chiral symmetry, 
we again expect an O($a^2$) scaling behaviour towards the continuum limit,
if the correlation functions are computed with the proper improved operators 
(see for example ref.~\cite{Bietenholz:2004wv}).

\vspace*{0.6cm}
\noindent {\bf Creutz fermions} 
\vspace*{0.6cm}

\noindent The Creutz-Dirac operator can be found in different forms in
ref.~\cite{Creutz:2007af,Borici:2007kz,Bedaque:2008xs}.
We have rescaled the Creutz-Dirac operator with a factor $R$ 
that we leave unspecified for the moment. 
As will be discussed below and in the appendix, 
this normalization factor $R$ is needed to obtain
at tree-level the correct continuum limit
for the correlation functions we have studied.
In the appendix we define all the relevant functions and we show that the 
Creutz operator can be brought to the form
\begin{equation}
D_{\rm C}(p) = i\, \sum_\mu \, \ppall_\mu\, \bar{\gamma}_\mu -
i\,\frac{a}{2}\sum_\mu \, \hat{p}_\mu^2\, \bar{\Gamma}_\mu  + m_0\,\mathbbm{1} ,
\label{eq:Cr_fin}
\end{equation}
where $\bar{\gamma}_\mu$ and $\bar{\Gamma}_\mu$ are linear combinations of
gamma matrices defined in the appendix.
We recall that Creutz fermions, as explicitly shown in the appendix, depend in
general on two paramenters $R$  and $C$.
From our final expression of the Creutz operator~\eqref{eq:Cr_fin}, we can
obtain the quark propagator for Creutz fermions.
In the continuum limit, the quark propagator can be written as
\begin{equation}\label{eq:contCreutz}
\widetilde{S}_{\rm C}^{cont} (p)=
\frac{- i\, \sum_\mu \, p_\mu\, \bar{\gamma}_\mu + m_0\,\mathbf{1}}
{\sum_\mu \sum_\rho \, p_\mu p_\mu\, \bar{a}_{\rho \mu } \bar{a}_{\rho \mu } + 
\sum_{\mu \ne \nu} \sum_\rho \, p_\mu p_\nu\, \bar{a}_{\rho \mu } \bar{a}_{\rho \nu } + m_0^2}
\end{equation}
which is not the continuum Dirac propagator unless the values of $C$ and $R$,
whose dependence is hidden in the matrices $\bar{a}$ and $\bar{\gamma}_\mu$, are chosen properly.\\

\noindent From equation~\eqref{eq:contCreutz} it is also clear that
in order to obtain the correct continuum limit of the \emph{quark propagator},
we have to impose that $\bar{\gamma}=\bar{a}^T\gamma$ is again a gamma matrix,
which is equivalent to say that the following relation holds true
\begin{equation}
\{ (\bar{a}^T\gamma)_\mu , (\bar{a}^T\gamma)_\nu \}
= \sum_{\rho,\sigma}\, \bar{a}_{\rho \mu}\,\bar{a}_{\sigma \nu}\,\{ \gamma_\rho , \gamma_\sigma \}
= 2\,\sum_\rho \, \bar{a}_{\rho \mu}\,\bar{a}_{\rho \nu}
= 2\,\delta_{\mu \nu} .
\end{equation}
Taking into account the form of the matrix $\bar{a}$ defined in the appendix, $\bar{\gamma}$ is a gamma matrix
if and only if the following two conditions hold
\begin{align}
\sum_\rho \, \bar{a}_{\rho \mu}\,\bar{a}_{\rho \nu} = 0 \mbox{ for } \mu \ne \nu
 &\Rightarrow \frac{3S}{C} = \pm 1 \Rightarrow \quad C = \frac{3}{\sqrt{10}}\label{eq:gamma1}\\
\sum_\rho \, \bar{a}_{\rho \mu}\,\bar{a}_{\rho \nu} = 1 \mbox{ for } \mu = \nu
 &\Rightarrow  \qquad \qquad \qquad R^2 = 3 + (\frac{3S}{C})^2\label{eq:gamma2}.
\end{align}
Therefore, from this discussion, it can be concluded that the right continuum limit of the
\emph{quark propagator} can be obtained only when $C=3/\sqrt{10}$ and $R=2$.\\

However,
motivated by the first version of \cite{Creutz:2007af},
we have decided to study as well the case $C=3/\sqrt{14}$. 
In this case
we know
that $\bar{\gamma}$ is not a gamma matrix, since the condition given by equation~\eqref{eq:gamma1}
does not hold. $R$ can be still determined and must be $R = 2\sqrt{2}$ in order to
satisfy equation~\eqref{eq:gamma2}. As expected, the analytical expression obtained for
the quark propagator in this case does not correspond to the continuum one.\\

But, as it will be shown in the next sections where we present our results,
the continuum limits for the pseudoscalar correlation function, mass and
decay constant turn out to be the correct ones for the two values of $C$
studied here.\\

There are two reasons to explain this behaviour.
On one hand, in the correlation function only the
contribution from the pole survives in the sum over the momenta. This implies that
all the possible crossed terms of the momentum which appear in the quark propagator,
whose origin is the non-realization of equation~\eqref{eq:gamma1}, are cancelled.
Moreover the properly chosen value of $R$ normalizes the quark propagator
in order to obtain the 'would-be' correct continuum limit in absence of
crossed terms.

\vspace*{0.6cm}
\noindent {\bf Borici fermions} 
\vspace*{0.6cm}

\noindent The action suggested by Borici~\cite{Borici:2007kz} is a slight 
modification
of the Creutz-Dirac operator and reads
\begin{equation}
\label{borici}
 D_B(p)=\sum_\mu i\gamma_\mu \ppall_\mu - i \frac{a}{2}\sum_\mu \Gamma_\mu \hat{p}_\mu^2 + m_0\,\mathbbm{1},
\end{equation} 
where $\Gamma_\mu$ are linear combinations of gamma matrices defined in the appendix.
The corresponding propagator can be written as
\begin{equation}
\label{prop}
 \widetilde{S}_{\rm B}(p)=\frac{-i\sum_\mu G_\mu(ap) \gamma_\mu + m_0\,\mathbbm{1}}{\sum_\mu G_\mu(ap)^2+m_0^2},
\end{equation} 
where the functions $G_\mu(ap)$, defined in the appendix, are trigonometric
functions of their argument.

\section{Correlation functions} 

In this section we give the expressions for the pseudoscalar and
proton correlation functions at tree-level of perturbation theory. 
While we evaluated the pseudoscalar correlation function for all 
lattice fermions considered in this work, we use the proton correlation 
function only to demonstrate how the O($a$) improvement works when the
Wilson average or equivalently the mass average is performed in case standard
Wilson fermions are considered. 

The interpolating fields describing the charged pions,
$\pi^{+}$ and $\pi^{-}$ are 
\begin{equation}
{\mathcal P}^{\pm}(x)\equiv {\mathcal P}^{1}(x)\mp i {\mathcal P}^{2}(x) 
\label{eq:ppm}
\end{equation}
where ${\mathcal P}^{a}(x)=\bar{\psi}(x)\gamma_{5}\frac{\tau^{a}}{2}\psi(x)$, with
$a=1,2,3$, is the
pseudoscalar density and $\tau^a$ are the standard Pauli matrices.
The computation for Wilson twisted mass is performed in the twisted basis, where
the form of the local operator~\eqref{eq:ppm} stays 
unchanged~\cite{Frezzotti:2000nk}.

The quark propagator can be decomposed in terms of the gamma matrices as
\begin{equation}
\tilde S(p)= S_{U}(p)\mathbbm{1} + \sum_{\mu}S_{\mu}(p)\gamma_{\mu} 
\end{equation}
in the case of overlap and Creutz fermions, while for twisted mass fermions
an additional term proportional to $\gamma_5$ is present
\begin{equation}
\tilde S(p)= S_{U}(p)\mathbbm{1} + \sum_{\mu}S_{\mu}(p)\gamma_{\mu} + S_{5}(p)\gamma_{5}. 
\end{equation}
With such a decomposition, the pseudoscalar correlation function can be 
written as\footnote{The Wtm propagator has also a flavour structure. For the pseudoscalar
correlation function only a single flavour component is needed.}
\begin{equation}
C(t)=\frac{N_c N_d}{L^3 T^2} \sum_{p_4} \sum_{p_4'} \sum_{\vec{p}}
 \sum_{\mu} e^{i(p_4-p_4')t} S_\mu(\vec{p},p_4) S_\mu^*(\vec{p},p_4'),
\label{eq:pscorrelator}
\end{equation}
with $\mu = U,1,2,3,4$ or $\mu = U,1,2,3,4,5$ depending on the kind of fermion
that is being considered.
$N_c$ is the number of colours and $N_d$ is the number of Dirac components. 
$T=aN_4$ and $N_4$ is the number of lattice points in the time direction.
The expression in eq.~(\ref{eq:pscorrelator}) can be evaluated as it stands, or, 
a time-momentum representation of the quark propagator can be obtained for a lattice with
infinite time extension, by performing the integration over $p_4$ analytically, 
see \cite{Thesis:2007} for a discussion in case of twisted mass fermions.
We have mostly used the representation of eq.~(\ref{eq:pscorrelator}), but checked 
for twisted mass fermions the results also in the time-momentum representation.
In the appendix we give the explicit expression for the Wilson twisted mass quark propagator 
in the time-momentum representation for a lattice with infinite time extent.

The local interpolating field describing the proton is
given by\footnote{The Greek (Latin) letters denote Dirac
(colour) components and $u$, $d$ denote the flavour content.
$C$ is the charge conjugation matrix
and $[\:]$ denotes spin trace.}
\begin{equation}\label{eq:defproton}
\mathscr{P}_{\alpha}(x)\equiv -\sqrt{2}\epsilon _{abc}\bigl[\bar{d}_{a}^{T}(x)\,C^{-1}
\gamma _{5} u_{b}(x)\bigr]\, u_{\alpha ,c}(x).
\end{equation}
The expression for the time dependence of the proton correlation function 
for standard Wilson fermions is then
\begin{equation}\label{eq:defproton2}
C_{\mathscr{P}\bar{\mathscr{P}}}(t)=\frac{N_{c}N_{d}}{L^{6}}\,\sum_{\vec{p}}\sum_{\vec{q}}\Bigl \{
L_{U}(\vec{p},\vec{q},t)+L_{4}(\vec{p},\vec{q},t)
\Bigr \},
\end{equation}
with the definitions
\begin{eqnarray}
L_{U}(\vec{p},\vec{q},t)&\equiv& S_{U}(-(\vec{p}+\vec{q}),t)
\Bigl\{(N_{d}+1)\,S_{U}(\vec{p},t)S_{U}(\vec{q},t)+\Bigr. \nonumber \\ 
&+& \Bigl.(N_{d}+3)\,\sum_{\mu =1}^{4}\,S_{\mu}(\vec{p},t)S_{\mu}(\vec{q},t)
\Bigr\}
\end{eqnarray}
\begin{eqnarray}
L_{\mu}(\vec{p},\vec{q},t)&\equiv& S_{\mu}(-(\vec{p}+\vec{q}),t)
\Bigl\{(N_{d}+3)\,S_{U}(\vec{p},t)S_{U}(\vec{q},t)+\Bigr. \nonumber \\ 
&+& \Bigl.(N_{d}+1)\,\sum_{\mu =1}^{4}\,S_{\mu}(\vec{p},t)S_{\mu}(\vec{q},t)
\Bigr\}.
\end{eqnarray}
$S_U$ and $S_\mu$ are the components of the quark propagator for standard Wilson fermions,
which can be obtained from the expression of the quark propagator for Wilson twisted mass fermions,
given in the appendix in eq.~\eqref{eq:propinftm1}, just by setting $\mu_q=0$.
These results have been cross-checked with a standard inverter on a 
cold gauge configuration.

\section{Scaling tests on correlation functions, masses and decay constants}

In this section we give some results from a scaling analysis in the lattice 
spacing. 
At tree-level a dimensionless quantity can be only a function
of $mL, a/L$ and $am$, where $m$ here indicates generically the quark mass.
To perform the continuum limit one can fix $mL$ to a certain value 
and the remaining dependence of the dimensionless quantity will be then 
in $a/L$. The continuum limit is then obtained sending $N=L/a$ to infinity.
In the following we will set $a=1$ and the $1/N$ and $1/N^2$ dependence of the 
dimensionless quantities under investigation will correspond to 
O($a$) and O($a^2$) scaling violations. We remind that if not infinity the time extent
$T$ will be always set to be proportianal to $L$.
 
In particular, we consider the correlation function $C$ at a fixed physical 
time $t/N$, the pseudoscalar decay constant $f_{\rm PS}$ and the 
pseudoscalar and proton masses $M$. 
This leads us to consider the dimensionless quantities
$N^3C(t/N)$, $NM$ and $Nf_{\rm PS}$.

We will start our discussion with an explicit 
demonstration of O($a$) improvement for
standard Wilson fermions from the 
Wilson and mass averaging procedures. We then turn over to the comparison 
of the size of scaling violations from all three lattice fermions considered 
here. In particular, we will compare the meson correlation function as well as
the corresponding meson mass and meson decay constant. 

\subsection{Wilson average and mass average for standard Wilson fermions }
\label{sec:wa}

\begin{figure}
\vspace{-1cm}
\hspace{-1.5cm}
\includegraphics[width=0.4\textwidth,angle=270]{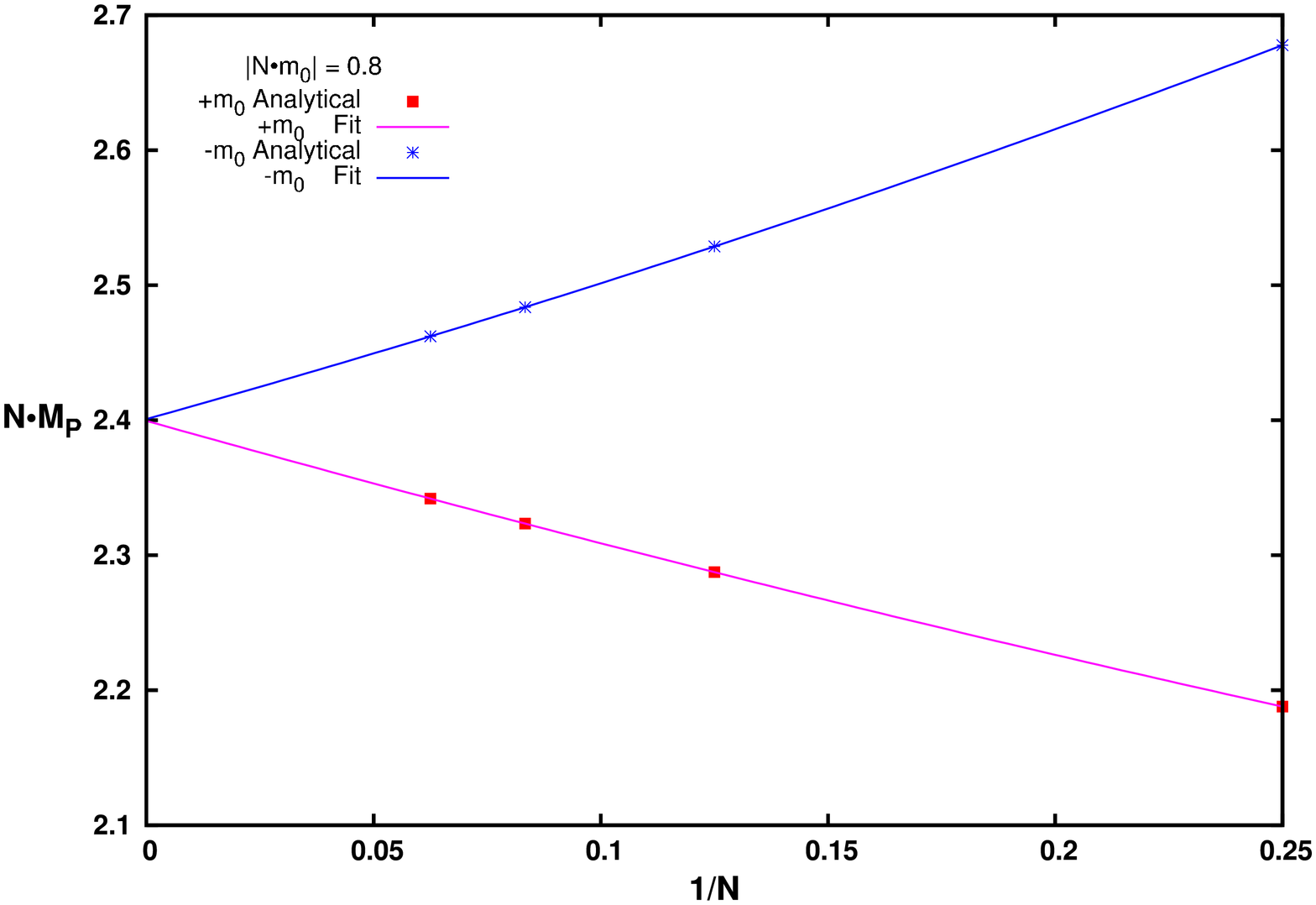}
\vspace{-1cm}
    \includegraphics[width=0.4\textwidth,angle=270]{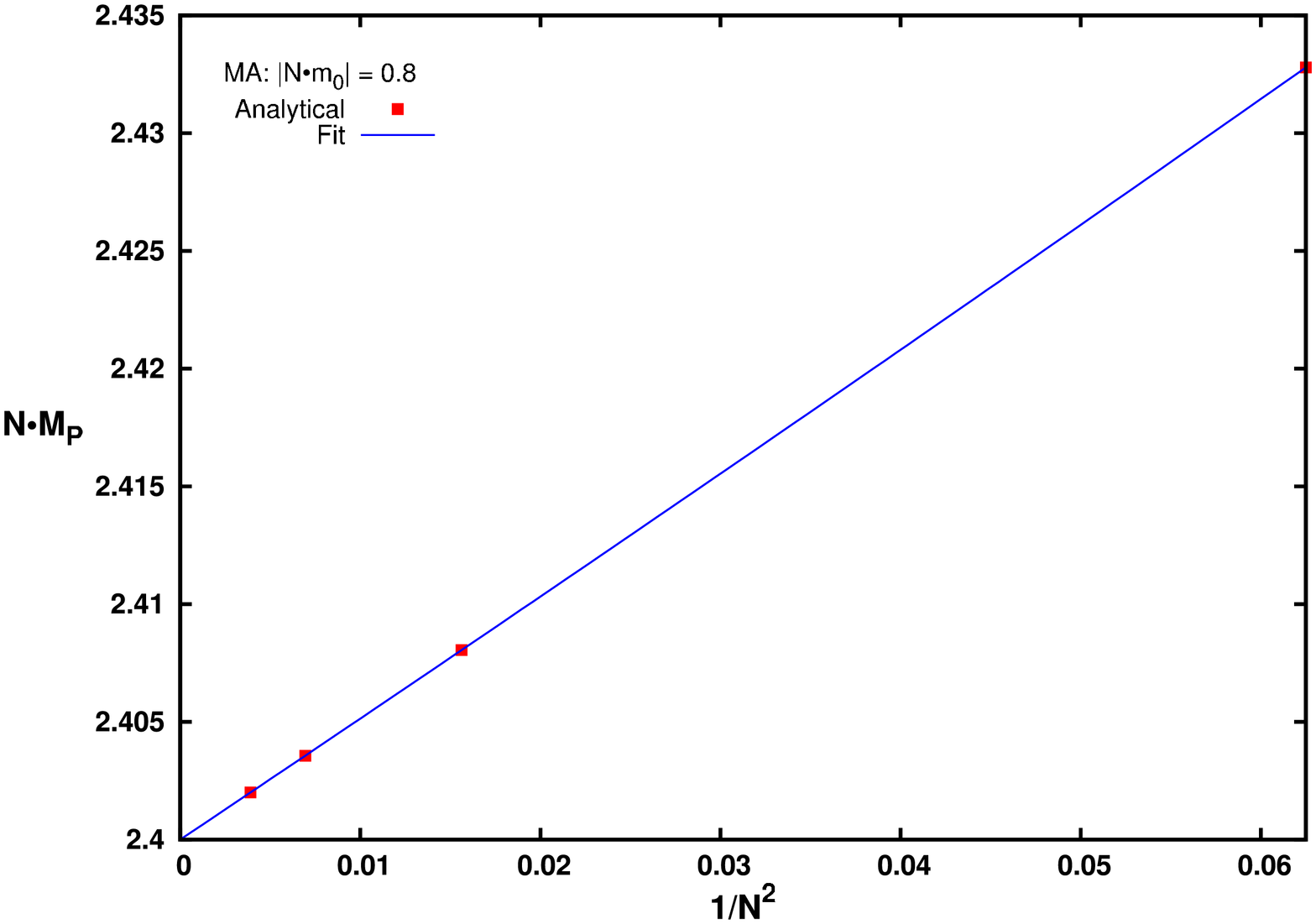}
\vspace{1cm}
\caption{In the left graph, the cutoff effects and
the continuum limit of the proton mass
obtained from two standard Wilson actions differing only
in the sign of the quark mass, $|Nm_{0}|=0.8$ are shown.
The lattices are $4\leq N\leq 20$.
The lines are fits according to eq.~(\protect{\ref{eq:wfit}}).
In the right graph, the average of the proton masses obtained from the same
two standard Wilson regularizations with quark masses $Nm_0=\pm 0.8$ (MA)
has been calculated. The solid line represents 
a fit to eq.~(\protect{\ref{eq:wfit}}).\label{fig:mapm1}}
\end{figure}

\noindent In Ref.~\cite{Frezzotti:2003ni} it has been demonstrated that
when averaging physical observables computed with standard Wilson actions, 
i.e. setting $\mu_q=0$ in eq.~(\ref{tmprop}), 
having opposite signs of the
quark mass $m_0$ (MA) or opposite signs of the
Wilson parameter $r$ (WA), these quantities are O($a$) 
improved\footnote{The MA is actually done taking into account the chirality of the 
correlation function under investigation~\cite{Frezzotti:2003ni}. This is
practically irrelevant for our study.}.
Since WA and MA at tree-level are equivalent, we will in the following only 
discuss the MA. As the physical observable we consider the proton 
mass as it can be obtained as the effective mass from the correlation 
function in eq.~(\ref{eq:defproton2}) using timeslices at
$t=4N$ and $t=4N+1$.  
For the computation we fix $|Nm_0|=0.8$. 
In the left graph of fig.~\ref{fig:mapm1} the behaviour of the proton mass $NM_{P}$
as a function of $\frac{1}{N}$ is given when two standard Wilson regularizations,
differing only in the sign of the quark mass, are used.

The behaviour of the proton mass in fig.~\ref{fig:mapm1} is 
linear in $1/N$ showing the expected O($a$) scaling violations of 
standard Wilson fermions.
The proton mass reaches, in both cases,
its continuum value of $NM_{P}=3|Nm_0|$,
as expected in the absence of interaction.
However, the slopes with which the continuum value is reached
i.e. the coefficients of the O($a$) cutoff effects differ in sign.
The right graph of fig.~\ref{fig:mapm1} shows the continuum approach of 
the proton mass when the proton mass is averaged 
over positive and negative values of the quark mass $Nm_0$. 
In this case, the proton mass is plotted against 
$1/N^2$ and the linearity in $1/N^2$ nicely shows the O($a$)-improvement when 
the MA is applied. 

In order to be more quantitative, 
we have used the following
fitting functions to describe the analytically computed values of the 
proton mass:
\begin{equation}\label{eq:wfit}
y_1=a_{0}+a_{1}\frac{1}{N}+a_{2}\frac{1}{N^{2}}
\; , \;\;
y_2=b_{0}+b_{1}\frac{1}{N^{2}}+b_{2}\frac{1}{N^{4}}.
\end{equation}
Here $y_1$ ($y_2$) is the physical observable under consideration
and its value in the continuum limit is given by the coefficient
$a_0$ ($b_0$).
We use two functional forms, the first formula of
equation~(\ref{eq:wfit}) for a leading $\frac{1}{N}$ behaviour (standard Wilson fermions)
and the second formula for O($a$)-improved quantities.\\

\noindent The two lines in the left graph of fig.~\ref{fig:mapm1}
originate from a fit to equation~(\ref{eq:wfit}) and
correspond to the proton mass
obtained from the same Wilson actions differing only in
the sign of the quark mass.
From the plot it is clear that in both cases
the value of the proton mass in the continuum limit is the same and
the expected one at tree-level of PT.

From the fit, the corresponding coefficients $a_1$ turn out to be
the same in magnitude but have
opposite signs for positive and negative quark masses.
Thus, performing the (MA), it is to be expected
that the O($a$) effects cancel and the scaling behaviour changes
from a $\frac{1}{N}$ to
a $\frac{1}{N^2}$ behaviour. This can indeed be seen in the right
graph of fig.~\ref{fig:mapm1}.
Inspecting the fit coefficients $a_2$ and $b_1$,
we find $a_{2}\approx b_{1}\approx 0.5$.
Therefore,
the magnitude of the leading order cutoff effects
does not only change from an O($a$) to an O($a^{2}$) behaviour but also
the O($a^{2}$) effects do not increase when performing the Wilson 
average with respect to the
standard case.

One interesting observation is the mass dependence
of the coefficients $a_1$ and $b_1$
of eqs.~\eqref{eq:wfit} for the pion and proton masses.
We observe that in the case of standard Wilson fermions, the coefficient $a_1/a_0$,
which determines the relative size
of the O($a$) cutoff effects, vanishes in the chiral limit proportionally
to $Nm_0$.
For Wtm fermions at maximal twist the coefficient $b_1/b_0$,
which determines the relative size
of the O($a^2$) cutoff effects, vanishes in the chiral limit, 
proportionally to $(N\mu_{\rm q})^2$.

\subsection{Comparing maximally twisted mass, overlap and Creutz fermions}

One interesting question is the {\em relative} size of cutoff effects when 
comparing maximally twisted mass, overlap and Creutz fermions. We have therefore 
performed a scaling analysis for the correlation functions themselves at a fixed 
physical distance, the pseudoscalar mass and the pseudoscalar decay constant.
Since all the quantities under investigation are O($a$)-improved, 
we show them all as a function of $1/N^2$. 

\begin{figure}
\vspace{0cm}
\hspace{2.5cm}
\includegraphics[width=0.4\textwidth,angle=270]{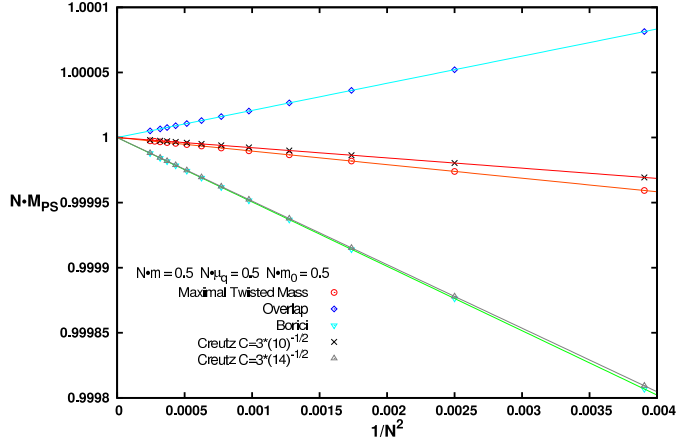}
\vspace{0.25cm}
\caption{The cutoff effects and
the continuum limit of the pseudo scalar mass is 
shown. We compare overlap, maximally twisted mass and Creutz fermions at
fixed quark mass, $Nm=0.5$, $N\mu_{q}=0.5$ and $Nm_{0}=0.5$, respectively.
The lattices are $4\leq N\leq 64$.\label{fig:allnm}}
\end{figure}

\begin{figure}
\vspace{-0.75cm}
\hspace{-1.5cm}
\includegraphics[width=0.4\textwidth,angle=270]{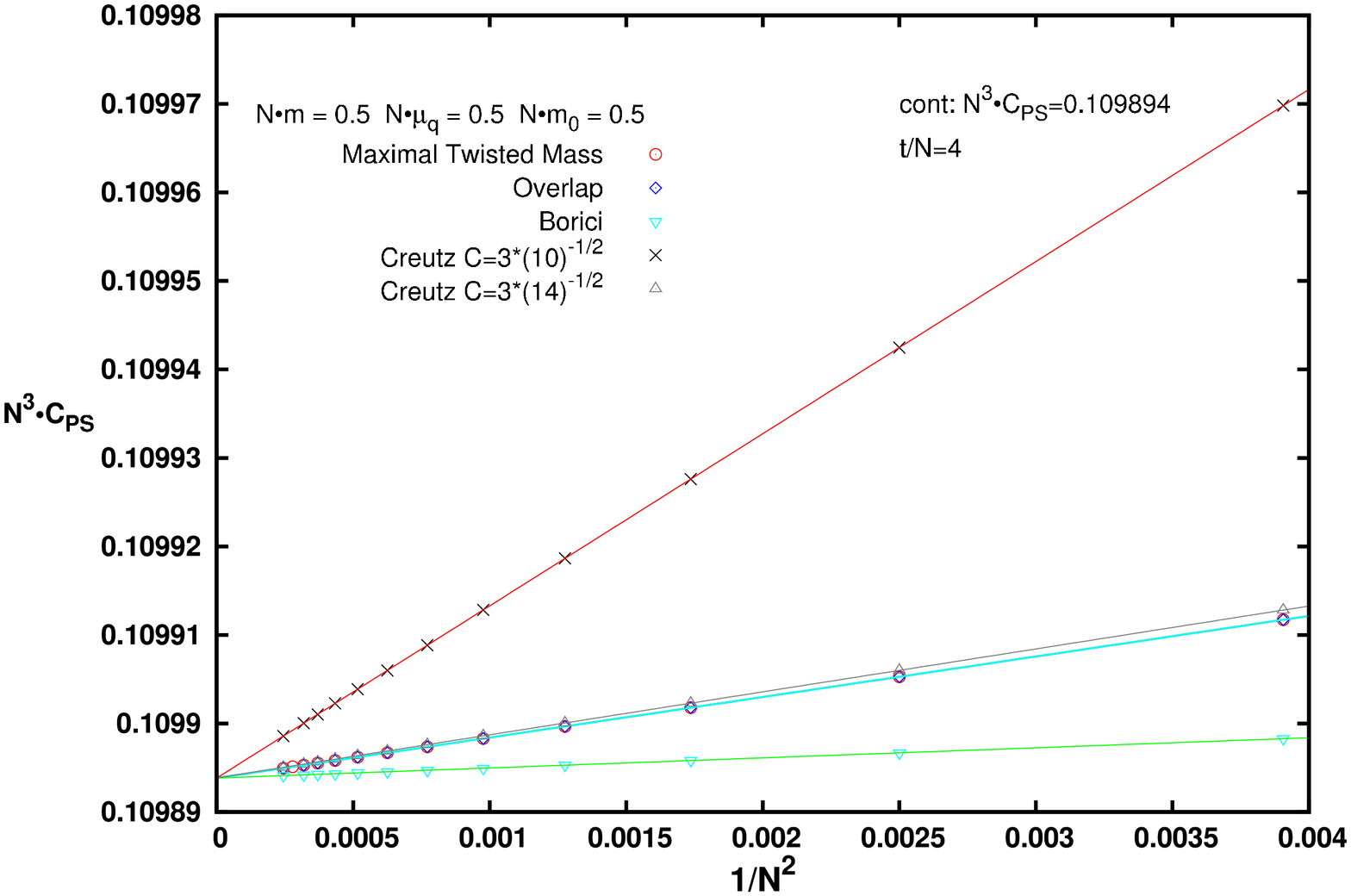}
\vspace{-1cm}
\includegraphics[width=0.4\textwidth,angle=270]{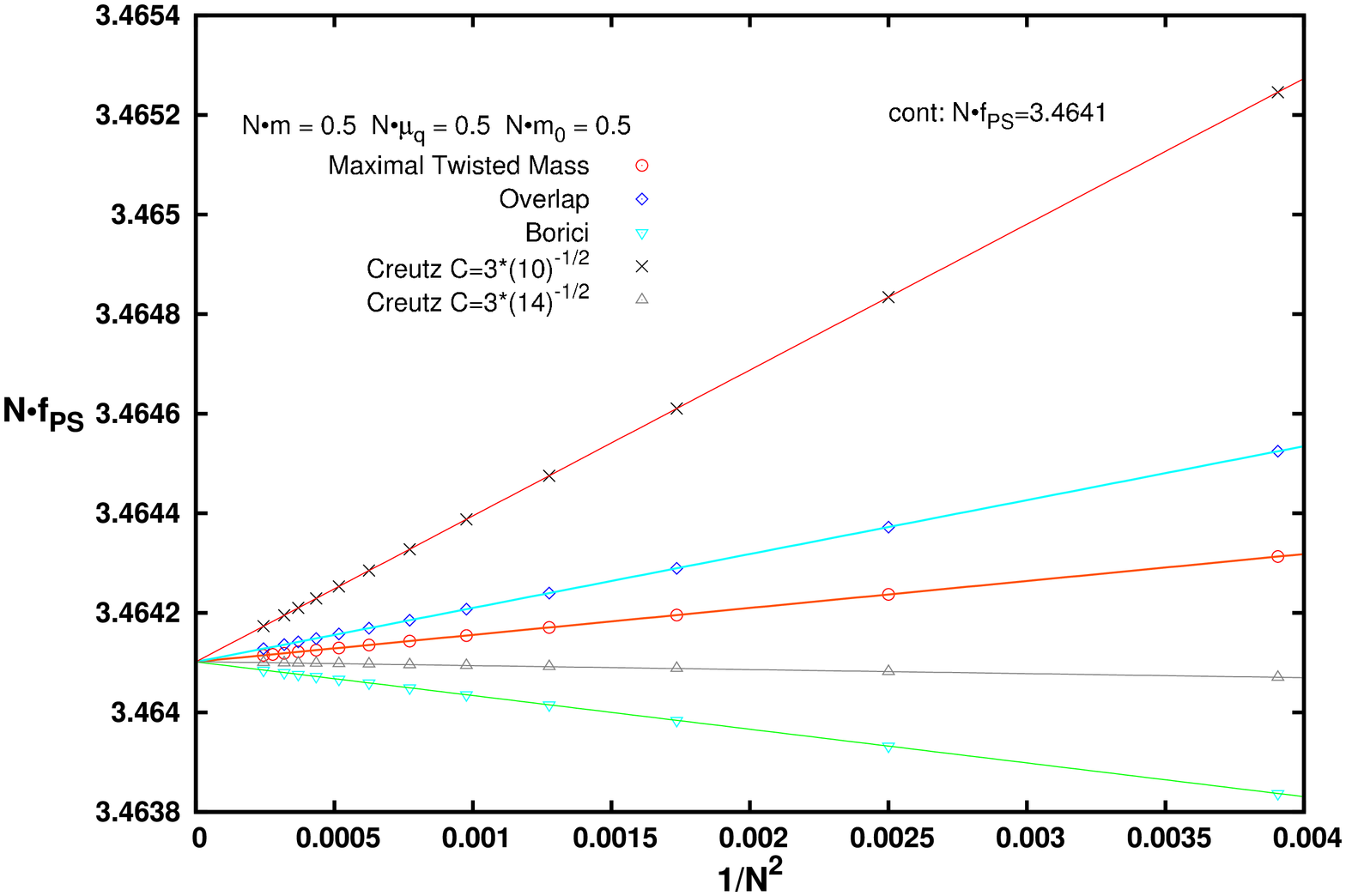}
\vspace{0.75cm}
\caption{In the left graph, the cutoff effects and
the continuum limit of the pseudo scalar correlation function is 
shown. We compare overlap, maximally twisted mass and Creutz fermions at
fixed quark mass, $Nm=0.5$, $N\mu_{q}=0.5$ and $Nm_{0}=0.5$, respectively.
The lattices are $4\leq N\leq 64$.
In the right graph, the pseudo scalar decay constants obtained from the same
lattice fermions as in the left graph
are shown.\label{fig:allcfn}}
\end{figure}

Let us start with the pseudoscalar mass
in  fig.~\ref{fig:allnm}.
As expected, we indeed find a nice linear 
behaviour of the mass as a function of $1/N^2$.
However, we observe that the different kind of lattice fermions we have studied show quite
distinct lattice artefacts at O($a^2$). 
This can also be seen
in fig.~\ref{fig:allcfn}, where we exhibit the 
dependence of the correlation function and the pseudoscalar 
decay constant as a function of $1/N^2$.

\begin{table}
\begin{center}
\begin{tabular}[c]{|c||c| r @{.} l | r @{.} l |}
\hline
   $NM_{\rm ps}$       & a  & \multicolumn{2}{|c|}{b}  & \multicolumn{2}{|c|}{c} \\
\hline
\hline
MTM       & $1 $ & $-0$&$0104167 $ & $ 0$&$000292154 $ \\
\hline
OVERLAP   & $1 $ & $ 0$&$0208333 $ & $ 0$&$000783943 $ \\
\hline
BORICI    & $1 $ & $-0$&$0494786 $ & $ 0$&$00558893  $ \\
\hline
CREUTZ - $\sqrt{10}$ & $1 $ & $-0$&$00781168 $  & $-0$&$010171$ \\
\hline
CREUTZ - $\sqrt{14}$ & $1 $ & $-0$&$0488288  $ &  $0$&$00287405  $ \\
\hline
\end{tabular}
\caption{Table of fit coefficients for the pseudo scalar mass using
$NM_{\rm PS} = a + b\frac{1}{N^2} + c\frac{1}{N^4}$.}
\label{tab:mps}
\end{center}
\end{table}

\begin{table}
\begin{center}
\begin{tabular}[c]{|c|| r @{.} l | r @{.} l | r @{.} l |}
\hline
    $Nf_{\rm PS}$       & \multicolumn{2}{|c|}{a} &  \multicolumn{2}{|c|}{b} &  \multicolumn{2}{|c|}{c}   \\
\hline
\hline
MTM       & $ 3$&$4641 $ & $ 0$&$0541266 $ & $ -0$&$000815548 $  \\
\hline
OVERLAP   & $ 3$&$4641 $ & $ 0$&$108253 $ & $  0$&$00554908 $ \\
\hline
BORICI    & $ 3$&$4641 $ & $  -0$&$0676637 $ & $ -0$&$00486739 $  \\
\hline
CREUTZ - $\sqrt{10}$ & $ 3$&$4641 $ & $ 0$&$293217   $ & $ -0$&$0770494 $  \\
\hline
CREUTZ - $\sqrt{14}$ & $ 3$&$4641 $ & $ -0$&$00790885 $ & $ -0$&$0367598  $ \\
\hline
\end{tabular}
\caption{Table of fit coefficients for the pseudo scalar decay constant using
$Nf_{\rm PS} = a + b\frac{1}{N^2} + c\frac{1}{N^4}$.}
\label{tab:fps}
\end{center}
\end{table}

\begin{table}
\begin{center}
\begin{tabular}[c]{|c|| r @{.} l | r @{.} l | r @{.} l |}
\hline
   $N^3C(t/N=4)$        & \multicolumn{2}{|c|}{a} &  \multicolumn{2}{|c|}{b} &  \multicolumn{2}{|c|}{c}  \\
\hline
\hline
MTM       & $ 0$&$109894 $ & $ 0$&$00457891 $ & $ -3$&$33302\cdot 10^{-5} $ \\
\hline
OVERLAP   & $ 0$&$109894 $ & $ 0$&$0045789 $ & $ 0$&$000181822 $ \\
\hline
BORICI    & $ 0$&$109894 $ & $ 0$&$00114427 $ & $ -0$&$00135812 $ \\
\hline
CREUTZ - $\sqrt{10}$ & $ 0$&$109894 $ & $ 0$&$0194625 $ & $ -0$&$00286602 $  \\
\hline
CREUTZ - $\sqrt{14}$ & $ 0$&$109894 $ & $ 0$&$00486428 $ & $ -0$&$002942 $ \\
\hline
\end{tabular}
\caption{Table of fit coefficient for the pseudo scalar correlation function
using 
$N^3C = a + b\frac{1}{N^2} + c\frac{1}{N^4}$.}
\label{tab:cps}
\end{center}
\end{table}

In order to see the cutoff effects in a quantitative way, the coefficients of the fit and the
corresponding fit function
for the three lattice quantities here considered,
obtained from the different regularizations
are presented in Table~\ref{tab:mps}, Table~\ref{tab:fps} and Table~\ref{tab:cps}, respectively.

From the results obtained here, no clear picture of a particularly good or bad 
fermion discretization emerges. While we find that indeed all three kind of
lattice fermions show the expected O($a$)-improvement, the (relative) size of the 
O($a^2$) effects depends pretty much 
on the observable considered. If at all, one could say 
that maximally twisted mass fermions show uniformly small O($a^2$) cutoff effects. 
On the other hand, it is somewhat amazing that Creutz fermions 
which break a number of important discrete symmetries do not suffer
from very large O($a^2$) cutoff effects. 
From our scaling analysis it is not possible to exclude a certain 
type of lattice fermion. Only scaling tests for the interacting theory 
will reveal the size of actual scaling violations of the observable considered. 
 
\section{Effects from non-optimal tuning}

This section is devoted to the question of effects when tuning is performed 
non-optimally. In particular, we study the cutoff effects when there is an 
O($a$) error in tuning to maximal twist.
As a second example, we consider 
the case when the quark masses of two lattice fermion formulations are not
exactly matched. This case is relevant for so-called mixed action simulations.

\subsection{Out of maximal twist}
\label{sec:omt}

\noindent Here we want to study a situation
when we allow an O($a$) error in
setting the untwisted quark mass to zero. In order to realize this
situation at tree-level of perturbation theory
we `force' these effects by simply
fixing the twisted mass to be the physical quark mass
and the untwisted mass is set to be proportional to $\frac{1}{N}$, as
$N\mu_{q}=\alpha$ and $Nm_{0}=\frac{\beta}{N}\backsim O(a)$
where $\alpha$ is kept fixed
and $\beta$ is a measure parametrizing the amount of violation
of the maximal twist setup. The
twist angle $\omega$ and the bare polar mass $M$ can be
obtained as a function of $\alpha$ and $\beta$ as
\begin{equation}\label{eq:womt}
\omega = \frac{\pi}{2}-\Bigl (\frac{\beta}{\alpha}\Bigr )\frac{1}{N}+ O(\frac{1}{N^2}), \qquad
NM= \alpha \Bigl
[ 1+\frac{1}{2}\Bigl (\frac{\beta}{\alpha}\Bigr )^{2}\frac{1}{N^{2}}+O(\frac{1}{N^4})\Bigr ].
\end{equation}
\noindent Therefore, even if the condition of maximal twist
can only be obtained up to O($a$) cutoff effects,
which is generally the case in practical numerical simulations,
the observables, which are only functions of the polar mass, are still
automatically O($a$) improved.\\

\begin{figure}
\vspace{-1cm}
\hspace{-1.5cm}
    \includegraphics[width=0.4\textwidth,angle=270]{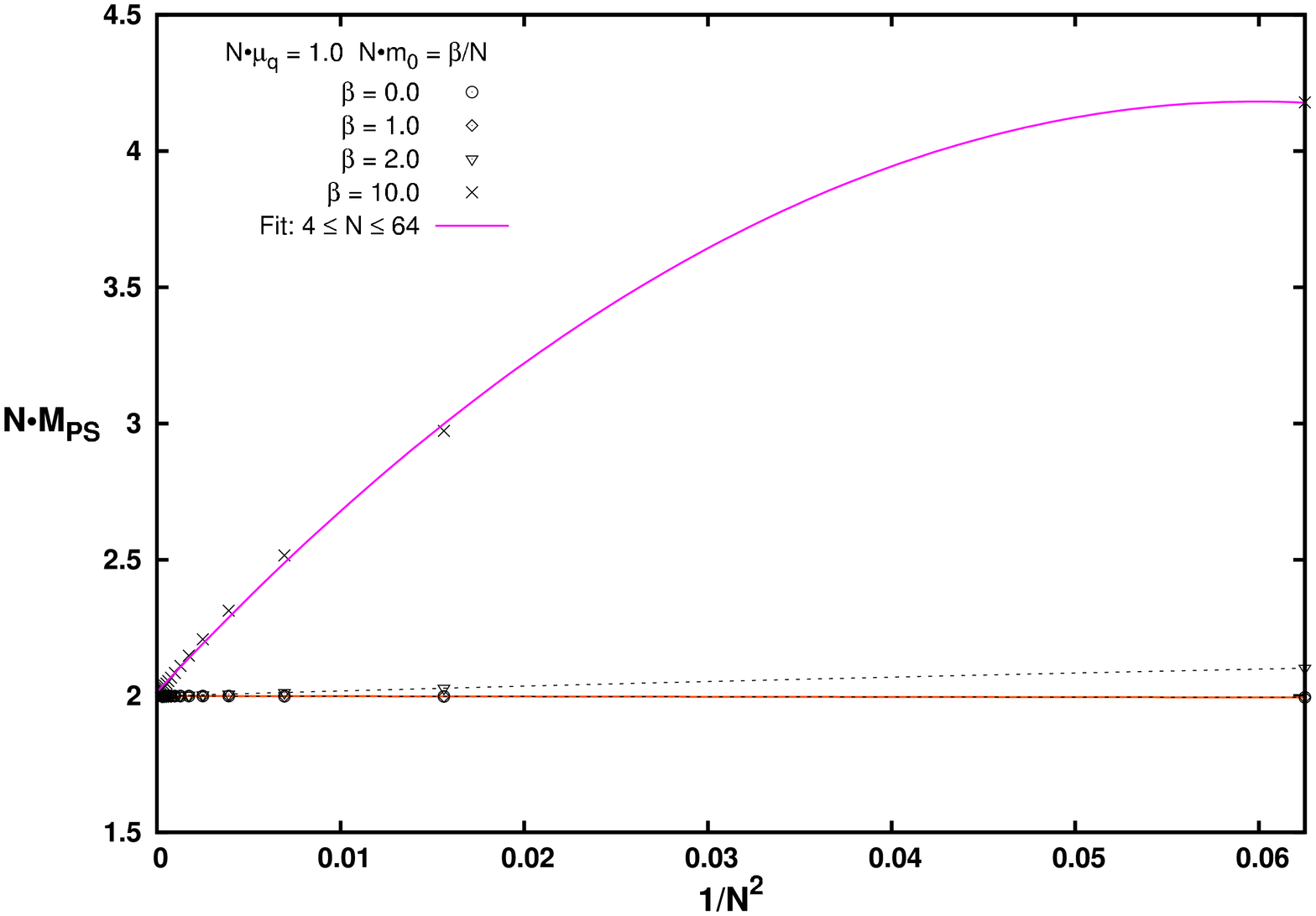}
\hspace{0.25cm}
\vspace{-1cm}
    \includegraphics[width=0.4\textwidth,angle=270]{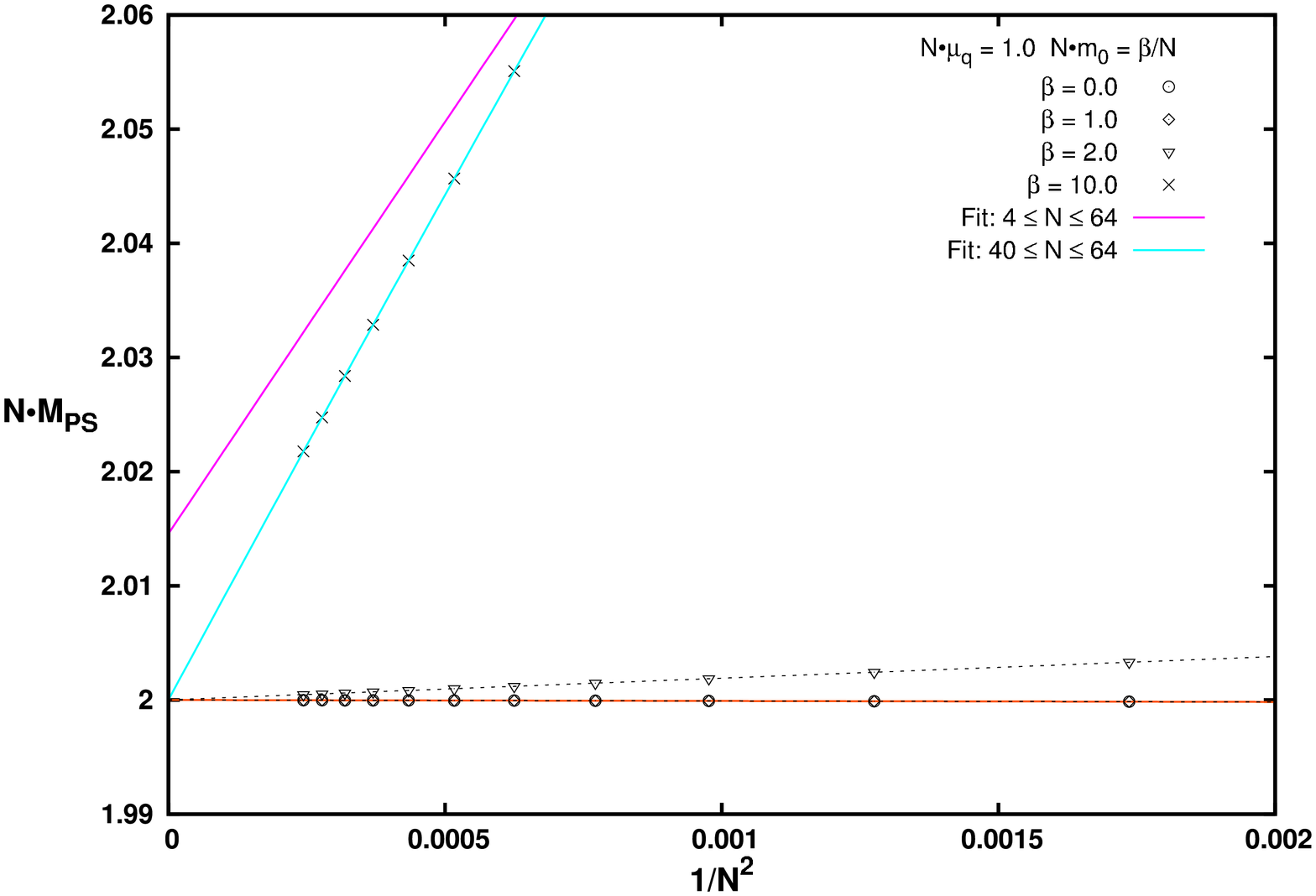}
\vspace{1.0cm}
\caption{Left graph: Behaviour of the pion mass as a function of $\frac{1}{N^{2}}$,
for lattices with size $4\leq N \leq 64$.
The twisted quark mass is set to $N\mu_{q}=1.0$ and the untwisted quark mass
is zero up to O($a$) cutoff effects i.e.
$Nm_{0}=\frac{\beta}{N}$ with $\beta = 0.0,1.0,2.0,10.0$.
Right graph: a zoom of the graph on the left with an additional fit for
the analytical data corresponding to $\beta =10.0$ which considers only large
lattices $40\leq N \leq 64$.\label{fig:outmt2}}
\end{figure}

\noindent Moreover,
equation~\eqref{eq:womt} also shows how the
size of the leading discretization effects
depends on the ratio between the untwisted
and twisted quark masses. This ratio in turn determines
the value of the lattice spacing at which the asymptotic $\frac{1}{N^2}$ scaling
sets in. Only when this ratio is small enough and hence the lattice
does not need to be chosen too large a
reliable continuum limit using reasonably sized lattices can be performed.
The left graph of fig.~\ref{fig:outmt2} demonstrates that
the asymptotic scaling sets in for lattices with $4\leq N \leq 64$
when $0\lesssim \frac{\beta}{\alpha}\lesssim 2$.\\

\noindent However, for $\frac{\beta}{\alpha}\,\gtrsim 10$
the continuum limit is not
reliable anymore if $N$ is chosen to be too small.
This can be seen in the right graph of fig.~\ref{fig:outmt2}.
Using only small values of $N$ leads
to an inconsistent continuum limit value.
Therefore, larger lattices are needed in order to obtain
the correct continuum behaviour as
can be also seen in the right graph of fig.~\ref{fig:outmt2}.
Here we have added a fit
of the data for a value of $\beta= 10.0$ taking only large lattices
into account,
i.e. using only values of $N \geq 40$.
In this case, indeed
the right continuum value is obtained.
Of course, for practical simulations, using only lattices with $N \gg 40 $ appears to be
rather unrealistic so to chose a correct ratio $\frac{\beta}{\alpha}$ becomes important.

\vspace{-0.5cm}

\subsection{Unmatched quark masses}

In this section we want to study
the continuum limit and the size of the cutoff effects of lattice quantities
constructed from ratios of physical observables computed on the lattice from two
different regularizations i.e. here Wilson twisted mass fermions at maximal twist
and overlap fermions. In particular, we want to study the situation when
both quark masses are not exactly fixed to the same value but differ
up to O($a^2$) effects.

The reason for studying 
such setup is that in real simulations using a mixed but 
O($a$)-improved action 
both masses can be fixed to the 
same value only up to O($a^2$)effects.
In order to realize non-matched quark masses,  
we fix the twisted 
quark mass exactly at $N\mu_q = 0.5$ and allow for an O($a^2$) error in setting the 
overlap quark mass,
\begin{equation}
Nm = 0.5 - v/N^2.
\end{equation}
We will vary the parameter $v$ from
$v=0$ to $v=4.0$.\\

\begin{figure}
\hspace{-1.5cm}
\includegraphics[width=0.4\textwidth,angle=270]{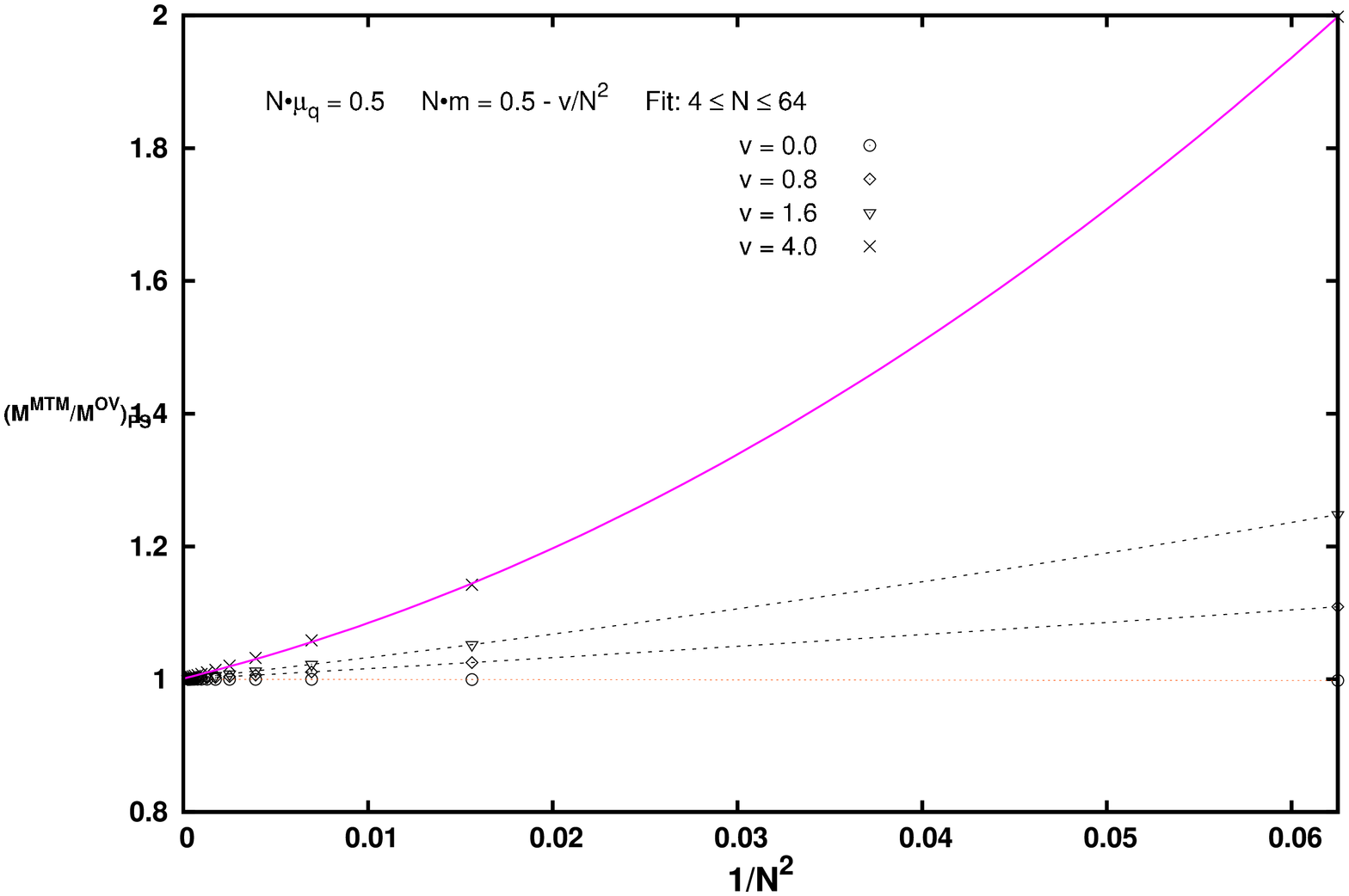}
\vspace{-1cm}
\includegraphics[width=0.4\textwidth,angle=270]{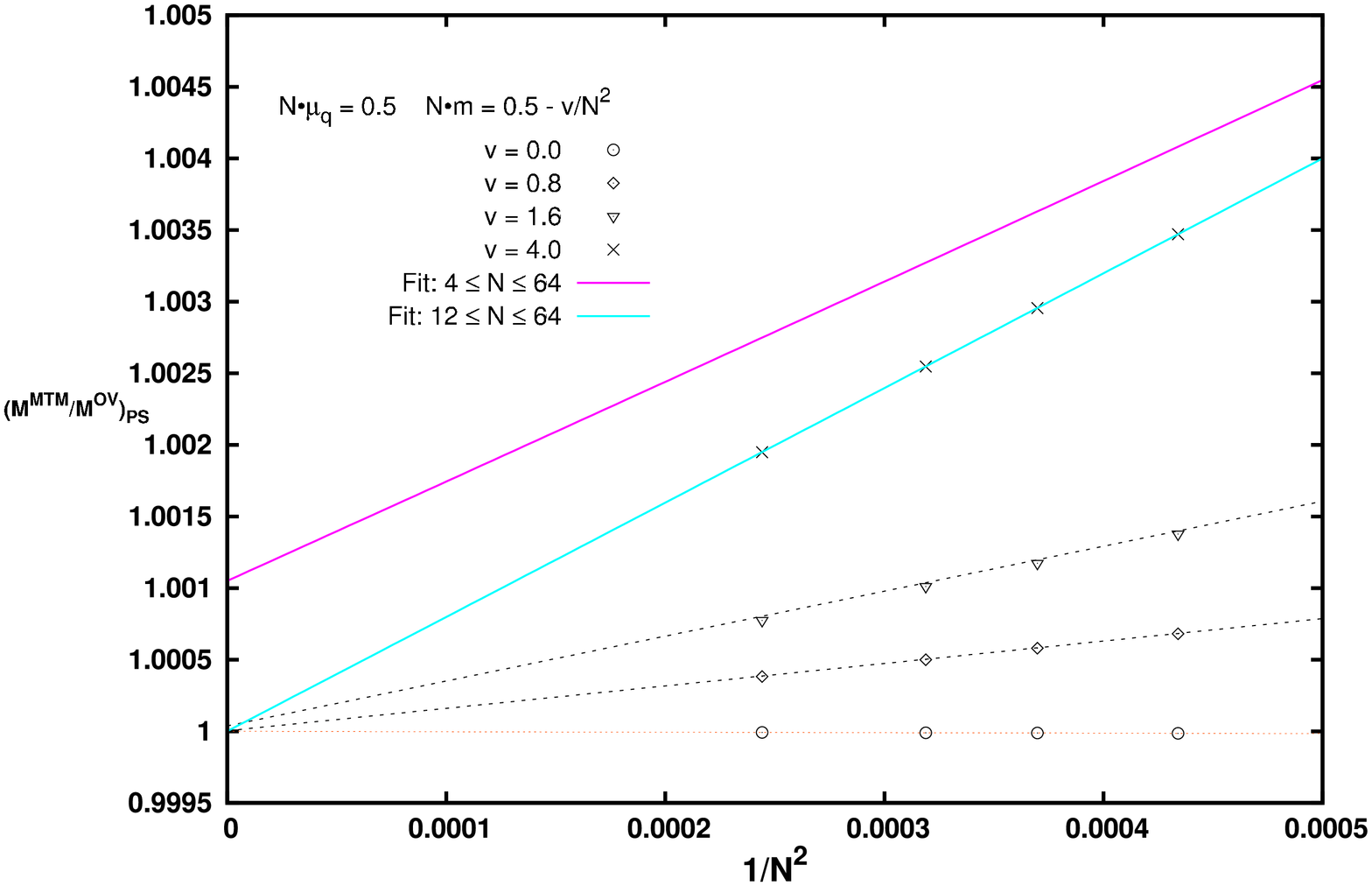}
\vspace{1cm}
\caption{The cutoff effects and continuum limit of  
the ratio of the pseudoscalar mass computed 
for maximally twisted mass and overlap fermions. 
In both graphs $N\mu_q=0.5$, $Nm=0.5-0.4/N^2$ and $t/N=4$. 
The left graph shows the full range of lattice sizes considered 
while the right graph represents a zoom. \label{fig:ov5}}
\end{figure}

Similar to the case of twisted mass at a non-optimally tuned value of the 
bare quark mass, the leading O($a^2$) cutoff effects can become very 
large when the quark masses are substantially mis-matched. 
This might even induce the danger of not achieving the right continuum 
limit when only too small lattice sizes are used. 

This is exemplified in fig.~\ref{fig:ov5}. In the left graph we show the 
scaling behaviour in $1/N^2$ for various values of the parameter $v$ for 
the full range of the lattice sizes.
If $v$ is small, we observe $1/N^2$ scaling violations.
Only in the case $v=4$, do the $1/N^4$ corrections become relevant. 
More quantitatively a linear fit in $1/N^2$ including all the data points
starting fron $N=4$ is reliable for $v \lesssim 1$.
While for $v \simeq 4$ we can include, in a linear fit in $1/N^2$,
only the data points starting from $N=12$.

Fig.~\ref{fig:ov5} demonstrates that not only the slope of 
the ratio of pseudoscalar masses versus $1/N^2$ increases as we make the value of $v$ 
larger, but also that the O($a^4$) effects become significant as can be seen 
from the curvature in the plot for $v=4$. Moreover, for $v=4$, the continuum 
limit fit is not reliable anymore if too small lattices are 
considered, which can be seen in the 
right graph of fig.~\ref{fig:ov5}.
Thus, if the matching is not performed with a good enough precision, 
large lattices have to be used to avoid higher order cutoff effects 
and perform a reliable continuum limit.

\section{Conclusions}

In this paper we have performed a scaling test in the lattice spacing
towards the continuum limit for three kinds of lattice fermions. 
Our setup has been tree-level of pertubation theory. 
The lattice fermions considered here were Wilson twisted mass, overlap 
and Creutz and Borici fermions. We looked at the pseudoscalar correlation function
at a fixed time and the corresponding pseudoscalar mass and decay constant. 

As a first step, we have verified automatic O($a$) improvement 
for Wilson twisted mass fermions and showed, with the example of the 
proton mass, the mechanisms of mass average. 

The relative comparison of all three kinds of lattice fermions we have considered
did not result in a clear picture in the sense that one lattice fermion
shows consistently smaller or bigger O($a^2$) lattice artefacts than the other. 
Rather we found that the sizes of O($a^2$) lattice artefacts
depend on the observable considered with perhaps the exception of 
maximally twisted mass fermions which shows a rather uniform behaviour 
with small O($a^2$) effects. Therefore, we expect that also in practical 
simulations the O($a^2$) lattice artefacts can turn out to be quite different
depending on the physical observable considered. 

Finally, we studied the situation when parameters are tuned 
non-optimally. We considered twisted mass fermions when an 
O($a$) error was allowed in the tuning to maximal twist. 
In addition, we looked at ratios of physical quantities built from overlap and twisted mass fermions.
In this case we allowed an O($a^2$) error in the matching of the quark masses. 
Our conclusion of these studies is that when the corresponding 
O($a^2$) error is too large, the continuum limit becomes not reliable 
if the lattice spacing is not small enough. 

\section*{Acknowledgements}

We thank M. Brinet, V. Drach and C. Urbach
for numerical checks of the Wilson twisted mass results presented here.
We thank Dru Renner for careful reading of this paper and useful comments.
A.S. thanks Chris Michael for discussions about Creutz fermions. A.S. acknowledges, 
for providing hospitality, DESY-Zeuthen where a big part of this work has been done.
J.G.L. thanks M. M\"uller-Preussker for discussions.
This work has been supported in part by  the DFG 
Sonder\-for\-schungs\-be\-reich/Transregio SFB/TR9-03.

\section*{Appendix}

In this appendix we summarize some of the formul{\ae} and definitions
which have been used in this work.

\vspace*{0.6cm}
\noindent {\bf Wilson twisted mass fermions} 
\vspace*{0.6cm}

For Wilson twisted mass fermions we give the analytic expression for the time-momentum
representation of the quark propagator on a lattice with infinite time extent.
Details of the computations can be found in ref.~\cite{Thesis:2007}.

The propagator reads\footnote{This expresion
is valid for all the possible values of
the discrete Euclidean time $t$, even if it is negative or zero.
This generalization has been done by using the
properties of the Fourier transform under change of sign of the
argument.},
\begin{equation}\label{eq:propinftm1}
\begin{split}
S \bigl( \vec{p},t \bigr)= \frac{1}{2\,\mathcal{E}_{1}}&\Bigl
\{\text{sgn}(t)\sinh E_{1} \gamma _{4}\mathbbm{1}_{f} -ia\mathcal{K}(\vec{p})\mathbbm{1}_{f}
+ \bigl[ \left( 1 - \cosh E_{1} \right) + aM(\vec{p})\bigr]\mathbbm{1}\mathbbm{1}_{f}\\
&- ia\mu _{q}\gamma _{5}\, \tau ^{3}\Bigr\}e^{- E_{1}|\frac{t}{a}|}\:
+\: \delta _{\frac{t}{a},0}\, \frac{1}{2\,(1 + aM(\vec{p}))}\mathbbm{1}\mathbbm{1}_{f}
\end{split}
\end{equation}
with\footnote{$\text{sgn}(t)$
is the sign of t,
and we have denoted $\text{sgn}(0)\equiv 0$.
It is just a convention in order to give one general
expression for the propagator for all possible values of $t$.}
\begin{equation}\label{eq:wilmass}
M(\vec{p})=m_{0}+\frac{a}{2}\sum_{i=1}^{3}\, \hat{p}_i^2 \qquad 
\mathcal{K}(\vec{p}) = \sum_{i=1}^{3}\, \gamma_{i} \ppall_i ,
\end{equation}
\begin{equation}
\cosh E_{1} = 1+ \frac{a^{2}\mathcal{K}^{2}(\vec{p}) + a^{2}M^{2}(\vec{p}) + a^{2}\mu _{q}^{2}}{2\,(1 + aM(\vec{p}))} ,
\end{equation}
\begin{equation}
\mathcal{E}_1 \equiv \left|1+aM(\vec{p})\right| \sinh E_1 .
\end{equation}

\vspace*{0.6cm}
\noindent {\bf Creutz fermions} 
\vspace*{0.6cm}

Our starting point for the analysis of the Creutz action is the operator
given by
\begin{equation}
D_{\rm C}(p) = i\, \sum_\mu \Big( \bar{s}_\mu(ap) +  \bar{c}_\mu(ap) \Big)\, \gamma_\mu  + m_0\,\mathbbm{1}
\label{creutzoperator}
\end{equation}
where $\mathbbm{1}$ is the identity matrix in Dirac space and the momentum has already been
reexpressed in terms of the 
pole as
\begin{equation}
q_\mu = \tilde{q} + p_\mu. 
\end{equation}
Starting from a reciprocal lattice labelled here by $q_\mu$ 
the poles of the Creutz-Dirac operator are localized at 
$(\tilde q,\tilde q,\tilde q,\tilde q)$ and 
$(-\tilde q,-\tilde q,-\tilde q,-\tilde q)$, where $\tilde q$ is related to the parameter 
$C$ by the equation $C=\cos(\tilde q)$. In the following, we consider only the former pole. 
In addition, we also define the parameter $S=\sin(\tilde q)$.

The relevant trigonometric functions for Creutz fermions are defined by
\begin{align}
\bar{s}_k(ap) &= \frac{1}{R}s_k(ap)\; , \;\;
\bar{s}_4(ap) = \frac{3S}{RC}s_4(ap)\\
\bar{c}_k(ap) &= \frac{S}{RC}c_k(ap)\; , \;\;
\bar{c}_4(ap) = \frac{3}{R}c_4(ap)
\end{align}
with the functions $s$ and $c$ given by
\begin{align}
s_1(ap) &= \left[ \ppall_1 + \ppall_2 - \ppall_3 - \ppall_4 \right]\\
s_2(ap) &= \left[ \ppall_1 - \ppall_2 - \ppall_3 + \ppall_4 \right]\\
s_3(ap) &= \left[ \ppall_1 - \ppall_2 + \ppall_3 - \ppall_4 \right]\\
s_4(ap) &= \left[- \ppall_1 - \ppall_2 - \ppall_3 - \ppall_4\right]\\
c_1(ap) &= -\frac{a}{2}\left[ \hat{p}_1^2 + \hat{p}_2^2 - \hat{p}_3^2 - \hat{p}_4^2 \right]\\
c_2(ap) &= -\frac{a}{2}\left[ \hat{p}_1^2 - \hat{p}_2^2 - \hat{p}_3^2 + \hat{p}_4^2 \right]\\
c_3(ap) &= -\frac{a}{2}\left[ \hat{p}_1^2 - \hat{p}_2^2 + \hat{p}_3^2 - \hat{p}_4^2 \right]\\
c_4(ap) &= -\frac{a}{2}\left[ \hat{p}_1^2 + \hat{p}_2^2 + \hat{p}_3^2 + \hat{p}_4^2 \right].
\end{align}

\noindent With the expression of eq.~(\ref{creutzoperator}) we obtain the 
corresponding propagator with standard manipulations as 
\begin{equation}
\widetilde{S}_{\rm C}(p) = 
\frac{- i\, \sum_\mu \Big( \bar{s}_\mu(ap) +  \bar{c}_\mu(ap) \Big)\, \gamma_\mu  + m_0\,\mathbbm{1}}
{\sum_\mu \Big( \bar{s}_\mu(ap) +  \bar{c}_\mu(ap) \Big)^2 + m_0^2}.
\label{creutzpropagator}
\end{equation}
In order to understand the continuum limit of the Creutz-Dirac operator and
the role of the factor $R$, we will  
use in the following the notation of Borici in \cite{Borici:2007kz}, with a scalar
product $(\gamma,x) \equiv \sum_\mu \gamma_\mu x_\mu$.\\
We thus write eq.~(\ref{creutzoperator}) as 
\begin{equation}
D_{\rm C}(p) = i\, (\gamma ,\bar{a}\tilde{s}(ap) + \bar{b}\tilde{c}(ap)) + m_0\,\mathbbm{1}
=  i\, (\bar{a}^T\,\gamma ,\tilde{s}(ap) + \bar{a}^{-1}\,\bar{b}\tilde{c}(ap)) + m_0\,\mathbbm{1} 
\end{equation}
where
\begin{equation}
\tilde{s} (ap) = \frac{1}{a}\,\left( \ppall_1,\ppall_2,\ppall_3,\ppall_4 \right)^T \qquad 
\tilde{c} (ap) = -\frac{a}{2}\left(\hat{p}_1^2,\hat{p}_2^2,\hat{p}_3^2,\hat{p}_4^2 \right)^T
\end{equation}
and the matrices $\bar{a}$ and $\bar{b}$ 
\begin{displaymath}
\bar{a} = \frac{1}{R}
\left(\begin{array}{cccc}
 1            &  1            & -1            & -1\\
 1            & -1            & -1            &  1\\
 1            & -1            &  1            & -1\\
-\frac{3S}{C} & -\frac{3S}{C} & -\frac{3S}{C} & -\frac{3S}{C}
\end{array}\right)
\end{displaymath}
\begin{displaymath}
\bar{b} = \frac{1}{R}\frac{S}{C}
\left(\begin{array}{cccc}
 1            &  1           & -1           & -1\\
 1            & -1           & -1           &  1\\
 1            & -1           &  1           & -1\\
 \frac{3C}{S} & \frac{3C}{S} & \frac{3C}{S} & \frac{3C}{S}
\end{array}\right).
\end{displaymath}

\noindent As a next step, we define
\begin{equation}
\bar{\alpha} \equiv \bar{a}^{-1}\,\bar{b}\; ,\;\; \bar{\gamma} \equiv \bar{a}^T\gamma
\; , \;\;
\bar{\Gamma} \equiv \bar{\alpha}\bar{\gamma} = \bar{\alpha}\bar{a}^T\gamma
\end{equation}
from which we obtain the final form of the Creutz-Dirac operator
\begin{equation}
D_{\rm C}(p) = i\, \sum_\mu \, \ppall_\mu\, \bar{\gamma}_\mu -
i \frac{a}{2}\sum_\mu \, \hat{p}_\mu^2 \bar{\Gamma}_\mu  + m_0\,\mathbbm{1}.
\end{equation}
This is the expression given in eq.~\eqref{eq:Cr_fin}.

\vspace*{0.6cm}
\noindent {\bf Borici fermions} 
\vspace*{0.6cm}

Here we give the definitions we have used for Borici fermions.
The $\Gamma$ matrices multiplying the Wilson term are defined by
\begin{equation}
\label{Gamma}
 \Gamma_\mu=\sum_\nu \alpha_{\mu\nu}\gamma_\nu,
\end{equation} 
with the definition 
\begin{equation}
\label{alpha_matrix}
 \alpha=\frac{1}{2}\left( \begin{array}{cccc}
1 & -1 & -1 & -1\\
-1 & 1 & -1 & -1\\
-1 & -1 & 1 & -1\\
-1 & -1 & -1 & 1
\end{array} \right).
\end{equation}
We incidentally note that $\sum_\mu \gamma_\mu = - \sum_\mu \Gamma_\mu$.

\noindent The trigonometric functions appearing in the Borici propagator are given by
\begin{align}
\label{def-g1}
 G_1(ap)&=\ppall_1 -\frac{a}{4}\left[ \hat{p}_1^2 + \hat{p}_2^2 - \hat{p}_3^2 - \hat{p}_4^2 \right]\\
\label{def-g2} 
 G_2(ap)&=\ppall_2 -\frac{a}{4}\left[ -\hat{p}_1^2 + \hat{p}_2^2 - \hat{p}_3^2 - \hat{p}_4^2 \right]\\
\label{def-g3} 
 G_3(ap)&=\ppall_3 -\frac{a}{4}\left[ -\hat{p}_1^2 - \hat{p}_2^2 + \hat{p}_3^2 - \hat{p}_4^2 \right]\\
\label{def-g4}
 G_4(ap)&=\ppall_4 - \frac{a}{4}\left[ -\hat{p}_1^2 - \hat{p}_2^2 - \hat{p}_3^2 + \hat{p}_4^2 \right].
\end{align}

\bibliographystyle{h-elsevier}
\bibliography{paper}

\begin{thebibliography}{10}

\bibitem{Cichy:2007vk}
K. Cichy et~al.,
\newblock (2007), arXiv:0710.2036 [hep-lat].

\bibitem{Frezzotti:2000nk}
ALPHA, R. Frezzotti et~al.,
\newblock JHEP 08 (2001) 058, hep-lat/0101001.

\bibitem{Frezzotti:2003ni}
R. Frezzotti and G.C. Rossi,
\newblock JHEP 08 (2004) 007, hep-lat/0306014.

\bibitem{Shindler:2007vp}
A. Shindler,
\newblock (2007), arXiv:0707.4093 [hep-lat].

\bibitem{Neuberger:1997fp}
H. Neuberger,
\newblock Phys. Lett. B417 (1998) 141, hep-lat/9707022.

\bibitem{Niedermayer:1998bi}
F. Niedermayer,
\newblock Nucl. Phys. Proc. Suppl. 73 (1999) 105, hep-lat/9810026.

\bibitem{Creutz:2007af}
M. Creutz,
\newblock (2007), arXiv:0712.1201 [hep-lat].

\bibitem{Borici:2007kz}
A. Borici,
\newblock (2007), arXiv:0712.4401 [hep-lat].

\bibitem{Bedaque:2008xs}
P.F. Bedaque et~al.,
\newblock (2008), arXiv:0801.3361 [hep-lat].

\bibitem{Luscher:1998pq}
M. L{\"u}scher,
\newblock Phys. Lett. B428 (1998) 342, hep-lat/9802011.

\bibitem{Hernandez:1998et}
P. Hernandez, K. Jansen and M. L{\"u}scher,
\newblock Nucl. Phys. B552 (1999) 363, hep-lat/9808010.

\bibitem{Carpenter:1984dd}
D.B. Carpenter and C.F. Baillie,
\newblock Nucl. Phys. B260 (1985) 103.

\bibitem{Thesis:2007}
J. Gonzalez~Lopez,
\newblock Cutoff effects for Wilson Twisted Mass Fermions at tree-level of
  Perturbation Theory, 2007.

\bibitem{Bietenholz:2004wv}
\xlf, W. Bietenholz et~al.,
\newblock JHEP 12 (2004) 044, hep-lat/0411001.

\end{thebibliography}

\end{document}